\documentclass[12pt]{article}
\usepackage{graphicx}
\usepackage{amsmath,amssymb,amsfonts}

\begin{document}

\title{Some remarks on the Bel-Robinson tensor and gravitational radiation}

\author{S. Hacyan
}

\renewcommand{\theequation}{\arabic{section}.\arabic{equation}}

\maketitle
\begin{center}

{\it  Instituto de F\'{\i}sica,} {\it Universidad Nacional Aut\'onoma de M\'exico,}

{\it A. P. 20-364, Cd. de M\'exico, 01000, Mexico.}

\end{center}
\vskip0.5cm

\begin{abstract}

The asymptotic form of the Bel-Robinson tensor in the gravitational radiation-zone is obtained in terms of the
mass quadrupole of the source. A comparison is made with the standard formula for the gravitational power
emission. The problem of a fully covariant description of gravitational radiation in terms of this tensor is
briefly discussed.
\end{abstract}

PACS: 04.30.Db; 04.30.Tv

Key words: gravitational waves

\section{Introduction}

Though the existence of gravitational waves is now well established, a manifestly covariant formulation of
the energy density and flux of gravitational radiation, produced by a physically realistic system, is still
lacking. The standard approach is based on the definition of a pseudo-tensor of energy-momentum \cite{ll,magg}.
This approach, however,  being coordinate dependent, has been the subject of decades of discussions (see
Kennefick's book\cite{ken} for an historical account). Finally, the discovery of the Hulse-Taylor pulsar\cite{ht}
came as a dramatic confirmation of the standard formulation.

From a theoretical point of view, the ``wave zone" of gravitational radiation can be defined covariantly, together
with its corresponding Poynting-vector \cite{adm}, but the problem of relating it in a fully covariant form to the
structure of its source, as in electromagnetism, remains open. An interesting approach is through the definition
of the Bel-Robinson tensor (Bel \cite{bel}), a fourth rank tensor that has been extensively studied by many
authors (see, e.g., references \cite{[1]} to \cite{[11]}).
In all cases, the underlying idea is that the Weyl tensor
$C_{\alpha\beta\gamma\delta}$ that describes the purely gravitational part of the space-time curvature is, in some
sense,  analogous to the electromagnetic field tensor $f_{\alpha\beta}$. Since the energy-momentum tensor of the
latter is a second rank tensor quadratic in $f_{\alpha\beta}$, the analogue of the energy-momentum tensor of the
former could be the Bel-Robinson tensor, which is quadratic in $C_{\alpha\beta\gamma\delta}$.
Following the original work of 
 Bel \cite{bel}, who proposed a definition of ``super-energy'' (and ``super-Poynting'')  for the
gravitational field, most of the previously cited authors attempted to relate this tensor to some appropriately
defined concept of energy. However, an application to concrete problems, such as the generation of gravitational
radiation by a binary pulsar, has not been achieved yet.

In the present paper, we calculate the asymptotic form of the Bel-Robinson tensor in the linear approximation of
general relativity, in order to shed some light on its physical significance and compare with standard results.
Since energy conservation is related to the existence of a time-like Killing vector, one can define a
``super-energy-momentum'' current in terms of this vector and the Bel-Robinson tensor, as in Ref.
\cite{[9],[7],[4]}. In analogy with the electromagnetic case, we deduce in Section 2 the explicit form of the 
super-Poynting vector in the asymptotic limit. In section 3, a comparison is made with 
electromagnetic radiation and the standard result for gravitational radiation. As briefly discussed in the last
section, there are some basic difficulties in relating the present approach to the well tested pseudo-tensor
formalism, basically due to the dimensions of the Bel-Robinson tensor.

\section{Basic equations}

In the weak field limit, the metric tensor is $g_{\mu \nu} = \eta_{\mu \nu} + h_{\mu \nu} $. Defining
$\overline{h}_{\mu \nu} \equiv h_{\mu \nu}- \frac{1}{2}\eta_{\mu \nu}  (\eta_{\rho \sigma} h^{\rho \sigma})$, the
metric is related to the energy-momentum tensor $T_{\mu \nu}$ through the equation\cite{ll,magg}
\begin{equation}
\Box \overline{h}_{\mu \nu} = -\frac{16 \pi G}{c^4} T_{\mu \nu}\label{dal},
\end{equation}
together with the gauge condition
\begin{equation}
\partial_{\mu} \overline{h}^{\mu \nu} = 0. \label{lor}
\end{equation}

In the radiation zone, at large distance $r$ from the source, the solution of Eq. \eqref{dal} has the asymptotic
form
\begin{equation}
\overline{h}_{ij} = -\frac{2 \pi G}{c^4} \omega^2  F(t, r) M_{ij},
\end{equation}
where
\begin{equation}
 F(t, r) \equiv  \frac{e^{-i\omega t + i kr}}{r}
\end{equation}
and the mass quadrupole is
\begin{equation}
-\omega^2 M_{ij} \equiv \frac{d^2}{dt^2} \int \rho (t,{\bf r}) x_i x_j ~dV = 2 \int T_{ij} dV
\end{equation}
(this last relation follows from the condition $\partial_{\mu} T^{\mu \nu} = 0$). A sinusoidal time dependence of
the source with frequency $\omega$ and wave number $k=\omega /c$ is assumed for simplicity.

The time-like components of $\overline{h}_{\alpha \beta}$ can be obtained from the gauge condition \eqref{lor}:
\begin{eqnarray}
-i k \bar{h}^{a0} + \partial_n \bar{h}^{an}&=&0 \\ \nonumber k^2 \bar{h}^{00} + \partial^2_{mn} \bar{h}^{mn}&=&0.
\end{eqnarray}

In the radiation zone, $kr \gg 1$, we have
\begin{equation}
\frac{\partial}{\partial x^{\alpha}}F(x^{\mu})= i k~ n_{\alpha} F(x^{\mu})~\Big(1+O(1/kr)\Big),
\end{equation}
where $n ^{\alpha}= (1, {\bf \hat{n}})$, and ${\bf \hat{n}} = {\bf r}/ r$. Thus we can set
\begin{equation}
\overline{h}_{\alpha \beta}=-\frac{2 \pi G}{c^4} \omega^2  M_{\alpha \beta} ({\bf \hat{n}}) F(t,r),
\end{equation}
in the understanding that
$$
M_{00} = M_{ij} n^i \hat{n}^j
$$
\begin{equation}
M_{0k} = -M_{kr} n^r.
\end{equation}
Notice, in particular, that
$
M_{\alpha \beta} n^{\beta}=0
$
and therefore
\begin{equation}
\partial_{\alpha} \bar{h}_{\beta\gamma} = ik ~n_{\alpha}\bar{h}_{\beta\gamma} \label{eso}.
\end{equation}

\subsection{Bel-Robinson tensor}

In the linear approximation of general relativity the Riemann tensor reduces quite generally to
\begin{equation}
R^{\alpha  \beta}_{~~\gamma \delta}= -2 \partial^{[\alpha} \partial_{[\gamma} h^{\beta]}_{~\delta]}.
\end{equation}

In vacuum, the Ricci tensor is identically zero and the Riemann tensor reduces to the Weyl tensor $C_{\alpha \beta
\gamma \delta}$. The  Bel-Robinson tensor $T_{\alpha \beta \gamma \delta}$  is defined as
\begin{equation}
T_{\alpha \beta \gamma \delta}=  C^{\mu ~~\nu}_{~\alpha ~~\gamma} C_{\mu \beta \nu \delta} + *C^{\mu
~~\nu}_{~\alpha ~~\gamma}~* C_{\mu \beta \nu \delta} ,\label{br}
\end{equation}
where $*C_{\alpha \beta \gamma \delta} = \frac{1}{2}\sqrt{-g} \epsilon_{\alpha \beta \mu\nu} C^{\mu \nu}_{~~~ \gamma \delta}$.  It is completely symmetric in its four indices, $T_{\alpha \beta \gamma \delta}= T_{(\alpha \beta \gamma
\delta)}$, traceless $T_{~\alpha \beta \gamma }^{\gamma}=0$, and divergence-free
\begin{equation}
\nabla^{\delta} T_{\alpha \beta \gamma \delta}=0
\end{equation}
in vacuum.

Using the condition \eqref{eso}, it follows from the definition \eqref{br}, with some lengthy but straightforward
algebra,
 \begin{equation}
T_{\alpha \beta \gamma \delta}=  \frac{\omega^8 }{2c^4} \Big( \frac{2\pi G}{c^4 r}\Big)^2 ~W ~n_{\alpha} n_{\beta}
n_{\gamma} n_{\delta} ,
\end{equation}
where
\begin{equation}
W = M^{\alpha}_{~\beta}M^{\beta}_{~\alpha}.
\end{equation}

Defining $\bar{M}=  M_{nn} $ and the trace-free quadrupole tensor $Q_{ij}=M_{ij} -\frac{1}{3} \delta_{ij}
\bar{M}$, we can express $W$ in terms of purely spatial components:
\begin{equation}
W=  (Q_{ij}\hat{n}^i \hat{n}^j)^2 -\frac{2}{3}\bar{M} Q_{ij}\hat{n}^i \hat{n}^j - 2
Q_{ik}Q_{jk}\hat{n}^i \hat{n}^j +Q_{ij}Q_{ij}+ \frac{2}{9}\bar{M}^2.\label{W}
\end{equation}

\section{From electromagnetism to gravitation}

Let us compare the electromagnetic and gravitational fields and see how the definition of the energy of the former
can be extended to the latter.

\subsection{Electromagnetic field}

In the dipole approximation, the electromagnetic field $f_{\alpha \beta} = 2\partial_{[\alpha} A_{\beta]}$ is
given en terms of the four-vector $A_{\alpha}$ whose space-like components, in the radiation zone $kr \gg 1$, are
\begin{equation}
{\bf A}=-ik ~{\bf p} F(t,r),
\end{equation}
with ${\bf p}$  the electric dipole (see, e.g., Jackson \cite{jac}). In this approximation, we can set $A_{\mu}=
-ik p_{\mu} F(t,r)$ in four-dimensional notation, where $p_{\mu}= ( p_0, {\bf p})$. Due to the Lorentz condition
$\partial_{\mu}A^{\mu}=0$, we have $n_{\mu} p^{\mu}= 0$ and therefore $p_0 ={\bf \hat{n}} \cdot {\bf p}$. It then
follows that  the electromagnetic energy-momentum tensor is
$$
T^{\alpha \beta}_{EM} = \frac{k^2}{4 \pi r^2} (p_{\mu}p^{\mu})~ n^{\alpha} n^{ \beta},
$$
with $p_{\mu}p^{\mu}=|{\bf \hat{n}} \times ({\bf \hat{n}} \times {\bf p})|^2$.

If the space-time admits a time-like Killing vector $\xi_{\alpha}$ such that $\nabla_{(\alpha} \xi_{\beta)}=0$,
the four-vector $J^{\alpha}= T^{\alpha \beta}_{EM}\xi_{\beta}$ is conserved if $\nabla_{\beta} T^{\alpha
\beta}_{EM}=0$, that is $\nabla_{\alpha} J^{\alpha}=0$. In Minkowski space-time, such Killing vector can be simply
$\xi^{\alpha}=(1,{\bf 0})$ and thus, in the radiation zone, $J^0$ is the energy density and ${\bf J}=cJ^0 {\bf
\hat{n}}$ is the Poynting vector. The electromagnetic power emitted by the dipole is $c\int J^0 r^2 d\Omega$.

\subsection{Gravitational field}

For the gravitational field in vacuum admitting a time-like Killing vector in the flat-space background, the
four-vector
$$
J_{\alpha} \equiv \lambda'~ T_{\alpha \beta \gamma \delta}~ \xi^{\beta} \xi^{\gamma} \xi^{\delta}
$$
is conserved, $\nabla_{\alpha} J^{\alpha}=0$, due to the properties of the Killing vector and the Bel-Robinson
tensor; a constant factor $\lambda'$ has been included for later convenience. Thus, with $\xi^{\alpha}=(1,{\bf
0})$, we can interpret $J^0$ as the ``super-energy'' density and ${\bf J}= c J^0 {\bf \hat{n}}$ as the
``super-Poynting vector''.

The angular integrals over products of the rectangular components of ${\bf \hat{n}}$ are
\begin{equation}
\int \hat{n}_i\hat{n}_j~ d\Omega= \frac{4 \pi}{3} \delta_{ij},
\end{equation}
\begin{equation}
\int \hat{n}_i\hat{n}_j \hat{n}_k\hat{n}_l ~d\Omega= \frac{4 \pi}{15} (\delta_{ij}\delta_{kl} + \delta_{ik}\delta_{jl}+\delta_{il}\delta_{jk}),
\end{equation}
from where it follows that
\begin{equation}
\int W ~d\Omega = 4\pi \Big(\frac{7}{15} Q_{ij} Q_{ij} + \frac{2}{9} \bar{M}^2 \Big).
\end{equation}
Accordingly the ``super-power'' $sP$ radiated is
\begin{equation}
sP =  \lambda \frac{G}{ c^5} \frac{\omega^8}{4\pi} \int W ~d\Omega ,
\end{equation}
redefining $\lambda' = \lambda ~c^6/8 \pi^3 G$ for comparison purposes. Observe that $sP/\lambda$ has dimensions
(energy / time$^3$).

\section{Discussion of results}

Compare the above formula for $sP$ with the well-known (and tested) standard formula for the total gravitational
radiated power \cite{ll,magg}:
\begin{equation}
P_{{\rm standard} }=\frac{G}{45c^5} \omega^6 Q_{ij}Q_{ij} ,
\end{equation}
with dimensions (energy / time) as it should be. The  difference (beside the trace $\bar{M}$) is that the
super-power is proportional to $\omega^8$ while the standard power is proportional to $\omega^6$. In order for the
two quantities to coincide (or at least be proportional), one could choose $\lambda \propto \omega^{-2}$, but then
the proportionality factor would not have a universal character since it would depend on the physical parameters
of each particular system. This problem with dimensions has been noticed by most previous authors (for instance, a
quantum of ``super-energy'' should be proportional to $\omega^3$ \cite{[12]}).

In order to further clarify this point, let us remind how an equivalent problem is treated in electromagnetism. A
distribution of electric charges and currents defines a four-vector $J^{\beta}_{matter}$, and the electromagnetic
energy-momentum tensor $T^{\alpha \beta}_{EM}$ is not conserved since
$$\nabla_{\beta} T^{\alpha \beta}_{EM}=c^{-1}f^{\alpha\beta}J^{\beta}_{matter}.$$ On the other hand, the charged particles producing the currents define an energy-momentum tensor
$T^{\alpha \beta}_{matter}$ of matter  such that
$$\nabla_{\beta} T^{\alpha \beta}_{matter}= -c^{-1}f^{\alpha\beta}J^{\beta}_{matter}$$
due to the Lorentz force on the particles (see Landau and Lifshitz \cite{ll}, Sect. 33). The net result is that
the total energy-momentum tensor, electromagnetic \emph{plus} matter, is conserved.

As for the  Bel-Robinson tensor, its divergence does not vanish in the presence of matter \cite{bel}. Accordingly,
in order to relate the emitted super-power to some mechanical properties of a physical system (such as a binary
pulsar), an independent definition of \emph{mechanical} super-energy would be required (for instance, for a
distribution of point-masses).  Such definitions for an electromagnetic field \cite{[1]} or a Klein-Gordon field has
been proposed in the past \cite{[9],[12]}. However, a useful definition should be obtained directly  from the \emph{dynamical}
equations
 of motion for massive particles,  analogous in general
relativity to the Lorentz force equation. As far as the writer knows, no such definition is known, and therefore a
fully covariant formalism based on the Bel-Robinson tensor and applicable to practical problems, such as
gravitational radiation, is still an open problem.


\end{document}